\documentclass[12pt,preprint]{aastex}

\usepackage{epsf}
\usepackage{dcolumn}
\usepackage{aastexug}

\newcommand{\BE}{\begin{equation}}
\newcommand{\EE}{\end{equation}}
\newcommand{\BA}{\begin{eqnarray}}
\newcommand{\EA}{\end{eqnarray}}

\def\be{\begin{equation}}
\def\ee{\end{equation}}
\def\bea{\begin{eqnarray}}
\def\eea{\end{eqnarray}}

\begin{document}

\renewcommand{\topfraction}{0.8}

\title{\Large\bf Lensing signals in the Hubble Ultra-Deep Field \\ using all 2nd order
shape deformations}

\shorttitle{Lensing using all 2nd order deformations}

\author{\bf John Irwin, Marina Shmakova,}

\affil{Stanford Linear Accelerator Center, Stanford University,
P.O. Box 4349, CA 94309, USA}

\author{\bf and Jay Anderson}

\affil{Physics and Astronomy Department, Rice University, MS-108,
Houston, TX 77705, USA}

\email{irwin@slac.stanford.edu, shmakova@slac.stanford.edu,
jay@eeyore.rice.edu}

\newpage

\begin{abstract}
The long exposure times of the \emph{HST} Ultra-Deep Field plus
the use of an empirically derived position-dependent PSF, have
enabled us to measure a cardioid/displacement distortion map
coefficient as well as improving upon the sextupole map
coefficient.  We confirmed that curved background galaxies are
clumped on the same angular scale as found in the \emph{HST} Deep
Field North. The new cardioid/displacement map coefficient is
strongly correlated to a product of the sextupole and quadrupole
coefficients. One would expect to see such a correlation from fits
to background galaxies with quadrupole and sextupole moments.
Events that depart from this correlation are expected to arise
from map coefficient changes due to lensing, and several galaxy
subsets selected using this criteria are indeed clumped.
\end{abstract}


\newpage

\section{Introduction} \label{intro}

This letter expands upon previous work of two of us
\citep{Irwin:2003qw, Irwin:2005nc}, hereafter [IS], aimed at
identifying lensing signals created by small-scale structure and
substructure, the ultimate purpose being the development of a
unique small-scale cosmological signal.

In [IS] we described and developed a non-linear lensing method
based on a complex power series lensing distortion map. The map is
derived from a symplectic coordinate transformation from a
reference frame at the telescope to a reference frame at the
source galaxy. It has a succinct complex power series form: $w_S =
w + a \bar w + b \bar w^2 + d_1 w \bar w + d_2 w^2$, where
$w=x+iy$ represents the position relative to the beam centroid.
This map is a generalization of the usual linear weak lensing
relationship. Each term in the series is a rotational eigenvector
and creates a distinct distortion of an azimuthally symmetric
source galaxy. There are two well-known linear terms: an
amplification (``convergence") and a term that produces a
quadrupolar distortion (``shear"). For lensing $a(=a_L)=-\gamma$.

\begin{figure}[h!]
\centering {\leavevmode
\includegraphics[width=4cm,height=4cm]{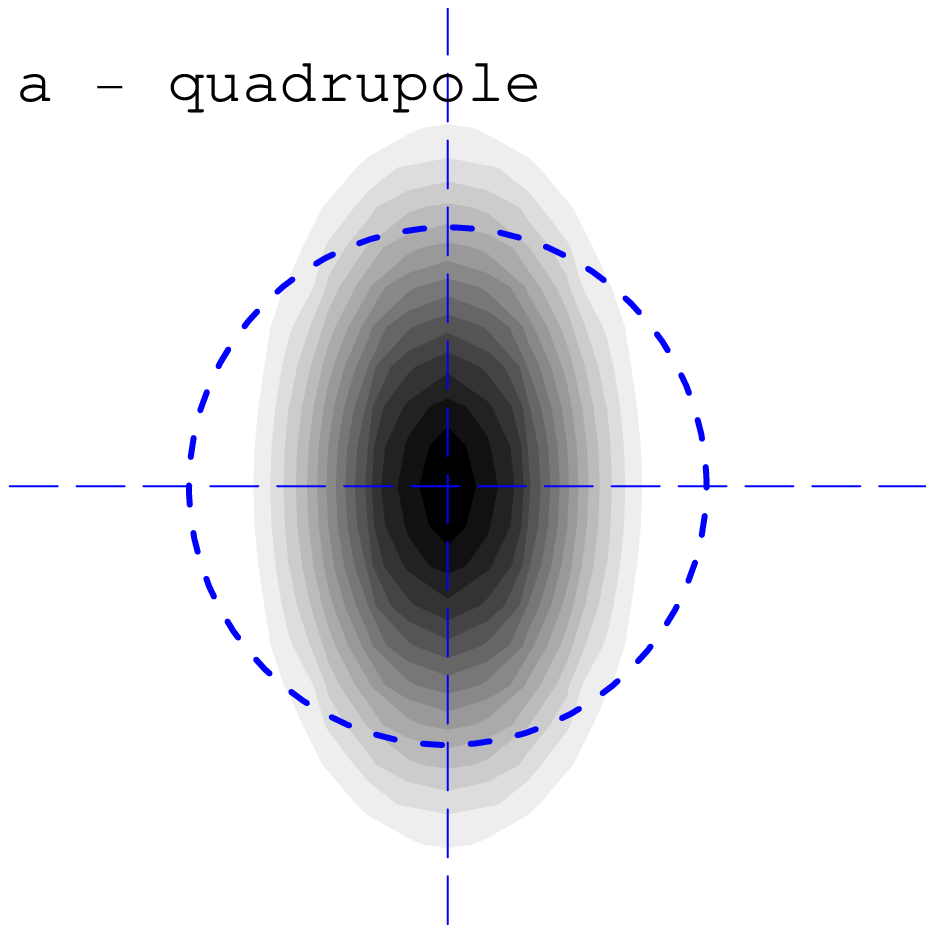}\,\,
\includegraphics[width=4cm,height=4cm]{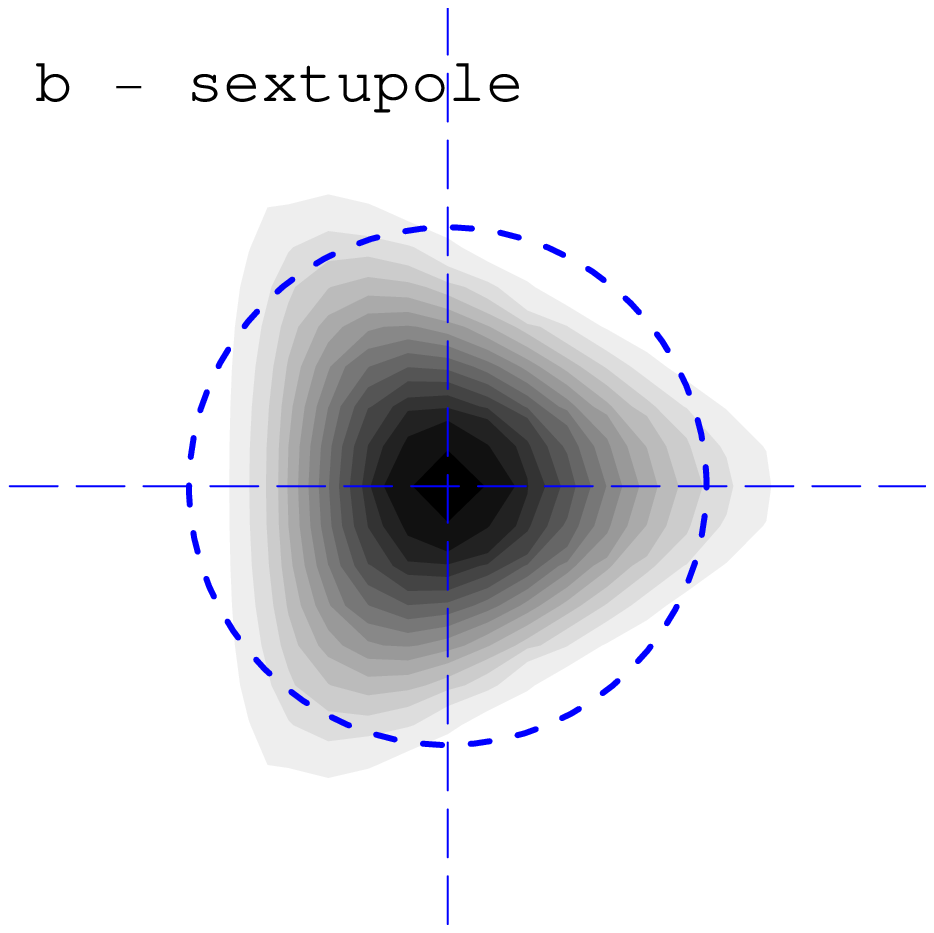}
\,\,
\includegraphics[width=4cm,height=4cm]{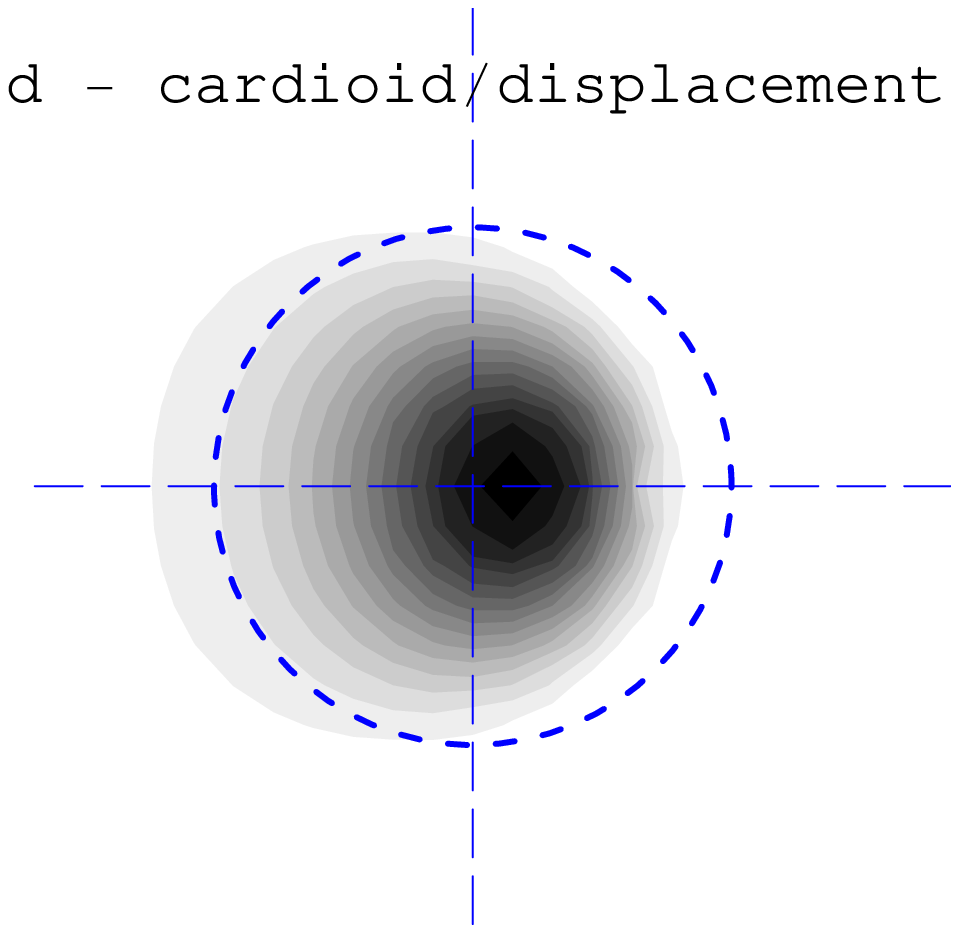}
} \caption[shapesDeform] {Shape deformations coming from different
terms of a map. The dashed circles correspond to an original
isophote.  These pictures demonstrate the deformation due to the
presence of single terms: either $a_{x}=0.3,\, b_{x}= -0.06,\,
d_{x}=0.02$.} \label{shapesDeform}
\end{figure}

The three 2nd-order map terms are: a sextupolar triangle-like
distortion, a cardioid distortion, and a quadratically increasing
displacement of circles.  Since lensing simultaneously produces
both the cardioid ($d_2$) and displacement ($d_1$) terms, they can
be denoted by a single coefficient ($d$), with $d_1=2d$ and
$d_2=\bar d$ . We call the so-combined $d$-terms the
``cardioid/displacement" term.  See fig.\ref{shapesDeform} for
examples of these shape distortions.\footnote{Other authors
(\cite{Bacon:2005qr,Goldberg:2004hh}) have referred to a related,
though differently parameterized, concept as Flexion I.}

The lensing distortion map embodies a distinction between
intrinsic galaxy shapes and allowed lensing distortions of that
shape.  The power of this approach will become particularly
evident here as we expand our consideration to include all
2nd-order map coefficients. The interrelationship of these
coefficients allows one to infer whether a particular background
galaxy is likely to have been lensed.

In [IS] the lensing distortion map formalism was applied to the
Hubble Deep Field North (HDFN). PSFs generated by \cite{TinyTim}
were convolved with mapped radial profiles. Parameters for the map
and the radial profile were simultaneously determined by a
least-squares fit to the background galaxy intensity matrices.
Curved background galaxies, defined by the relative orientation
(for that galaxy) of the quadrupole and sextupole map
coefficients, were found to be spatially clumped when compared to
randomly selected background galaxy subsets, suggesting that some
fraction of curved galaxies are the result of lensing by
foreground haloes and/or subhaloes.  We find it improbable that
the observed clumping could arise from intrinsic alignment of
background galaxies since the members of identified clumps have a
wide range of z-values.

We report here the results of extending that analysis to the
\emph{HST} Ultra-Deep Field (UDF).  Due to the longer UDF exposure
and access to an empirically derived time-averaged field-dependent
PSF for the Hubble wide-field camera (WFC)\citep{Anderson:2006},
we have again observed the clumping of curved galaxies and we have
successfully measured an additional second-order map coefficient
representing the cardioid/displacement distortion.

We find a strong correlation between the the cardioid/displacement
coefficient and a product of the quadrupole and sextupole map
coefficients. Such a correlation is expected from a fit to
background galaxy shapes. Galaxies that do not comply with this
correlation may be considered as candidates for quadrupole or
sextupole lensing or both. We present results supporting this
hypothesis.

\section{UDF data processing} \label{data}

The F606W UDF data set which we analyze in this paper consists of
112 half-orbit exposures. Each exposure was taken at a different
pointing, with two primary orientations that were rotated by
$\approx 90^\circ$. We transformed each frame of the field into a
distortion-free frame which one of us \citep{Anderson:2005} had
constructed for the WFC.

Using the distortion-free transform of the first UDF frame as a
frame of reference and bright point-like galaxies as locaters, we
next determined linear transformations between the distortion-free
transform of each of the remaining frames and this reference
frame. These transformations were then used to map the center of
each pixel in each frame into the reference frame.  We removed the
sky from the images by examining pixels in 800x800 patches and
setting the background to zero.

The construction of the super-image is an iterative process. The
super-image is chosen to be 2x super-sampled with respect to the
image pixels.  To construct this 8200x8200 super-image, we proceed
pixel-by-pixel finding the closest corresponding pixel in each
individual frame.  This provides 112 independent estimates of the
flux for each pixel. Since the individual frames have slightly
different exposure times, we divide the flux by the ratio of the
image's exposure to a reference-image exposure time of 1200
seconds.  We employ simple sigma-clipped averaging to find a
representative value for that pixel, obtaining a first super-image
estimate. Only one out of four pixels is statistically
independent.

So that the super image can be interpolated without high frequency
noise we smooth the first composite image with the kernel
$$
\left[\matrix{0.0625 & 0.1250 & 0.0625 \cr
                 0.1250 & 0.2500 & 0.1250 \cr
                 0.0625 & 0.1250 & 0.0625}\right].
$$
\noindent Then for each pixel in each individual exposure, we find
the difference between it and the interpolated super-image (from
the first iteration), thereby obtaining 112 frames of residuals.
We add the sigma-clipped average of these residuals to the
original pixel value, to get the final value of each super-image
pixel.

To ensure that the galaxy shapes we are measuring are intrinsic to
the galaxies and not artifacts of the detector, we propagate the
model PSFs through the same stacking procedure as used for the
galaxy images. To get a PSF appropriate for our super images, we
generated an array of artificial stars in our reference frame. The
stars had total fluxes of $10^5$ photons and were separated by
$\sim 100$ pixels and placed with pixel phases of $(0,0)$,
$(0.5,0)$, $(0,0.5)$ and $(0.5,0.5)$.  We used the inverse
transformations to determine where each of these stars landed in
each of the 112 exposures, and then used the spatially variable
PSF model of \cite{Anderson:2006} to generate an artificial set of
exposures with this grid of stars in them.

 Finally, we combine together the
PSFs at different pixel phases to get a x4 representation of the
PSF. It is this x4 PSF which we convolve with a x4 mapped radial
profile to find the best fit to each galaxy image.

Since the fit procedure to find map parameters is designed to
analyze galaxies with a single dominant peak, we have developed
software to identify and select such galaxies. This software
identifies all peaks within each galaxy (previously selected using
SExtractor software\cite{sextractor}) and based on criteria such
as separation, flux, and pixel height, decides whether these peaks
should be considered part of a single peak. After all peaks are
processed then, based upon a second set of criteria such as total
flux and maximum-to-minimum height ratio, satisfactory
central-peak regions are chosen to represent selected galaxies.

The fit-method is the same as that described in Section 3.3 of
[IS]. The radial profile is taken to be a quadratic polynomial in
$r^2$ times a Gaussian.  A condition is imposed that the
polynomial be positive. The map has four pairs of parameters: the
centroid, the real and imaginary quadrupole coefficients, the real
and imaginary sextupole coefficients, and the real and imaginary
cardioid/displacement coefficients. The procedure begins by
determination of the full-width-half-maximum of the image and its
ellipticity. Then, without yet introducing the PSF,  the radial
profile and map coefficients are fit using a least-square
difference with the selected region of the galaxy intensity
matrix. Convolutions with ever larger footprint PSFs are similarly
fit until the final PSF is convolved with the mapped radial
profile and fit.

\begin{figure}[h!]
\centering
{\leavevmode\includegraphics[width=2in,height=4cm]{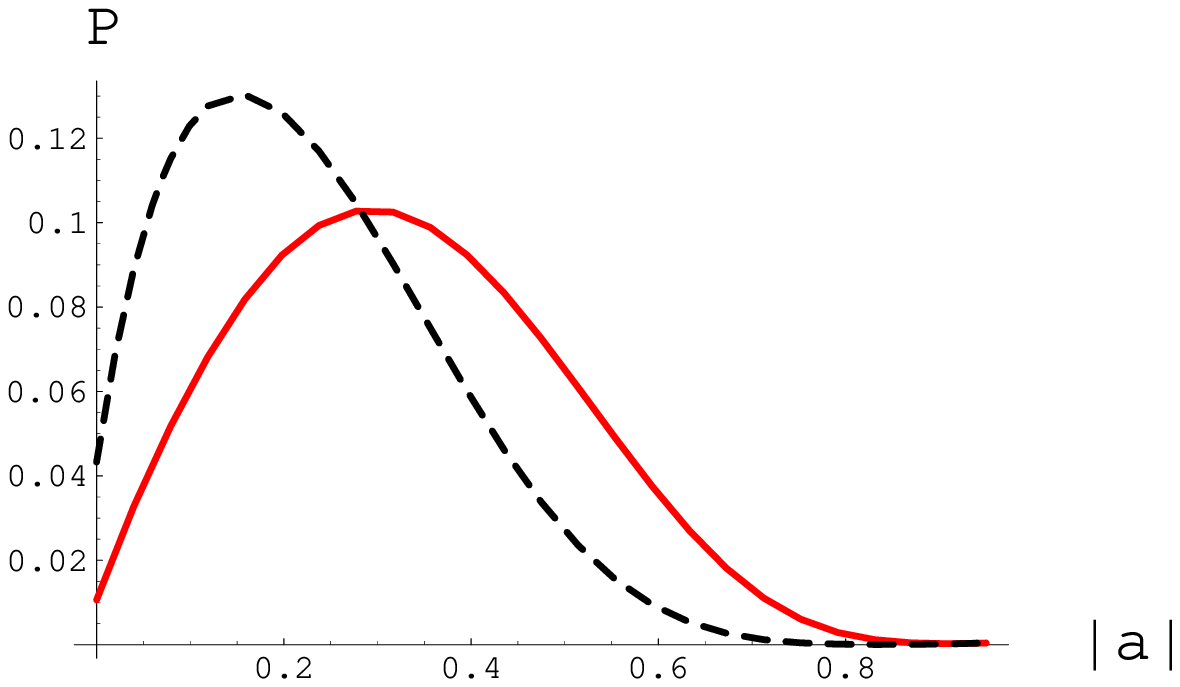}\,\,
\includegraphics[width=2in,height=4cm]{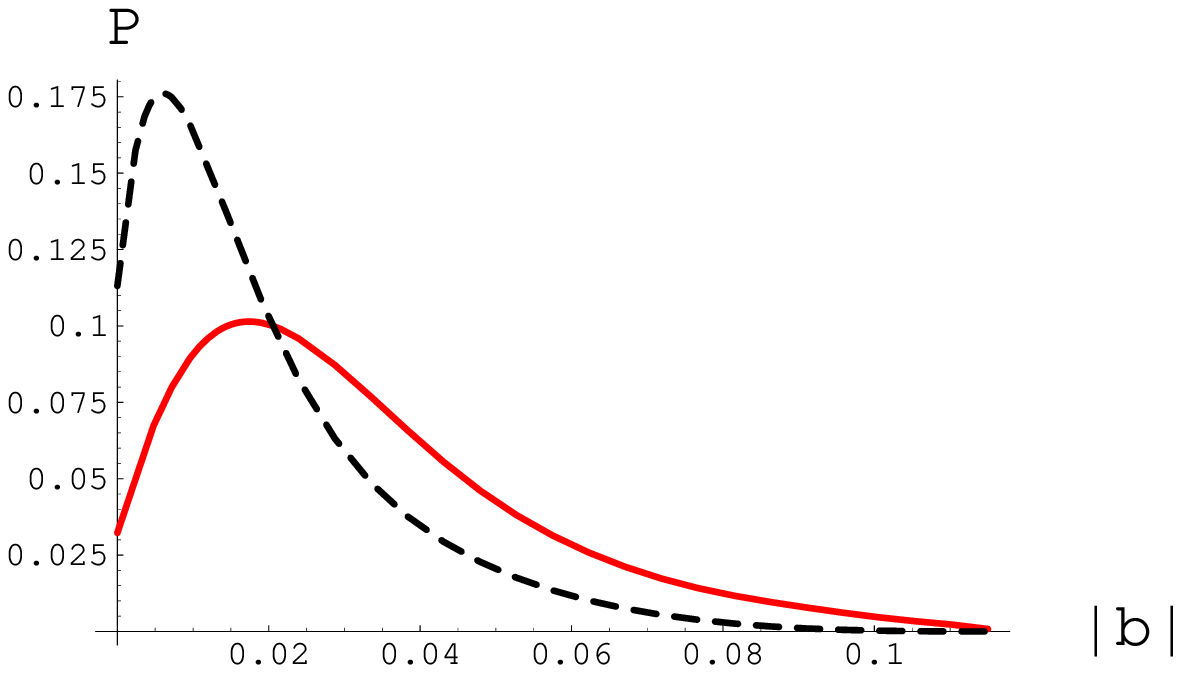}\,\,
\includegraphics[width=2in,height=4cm]{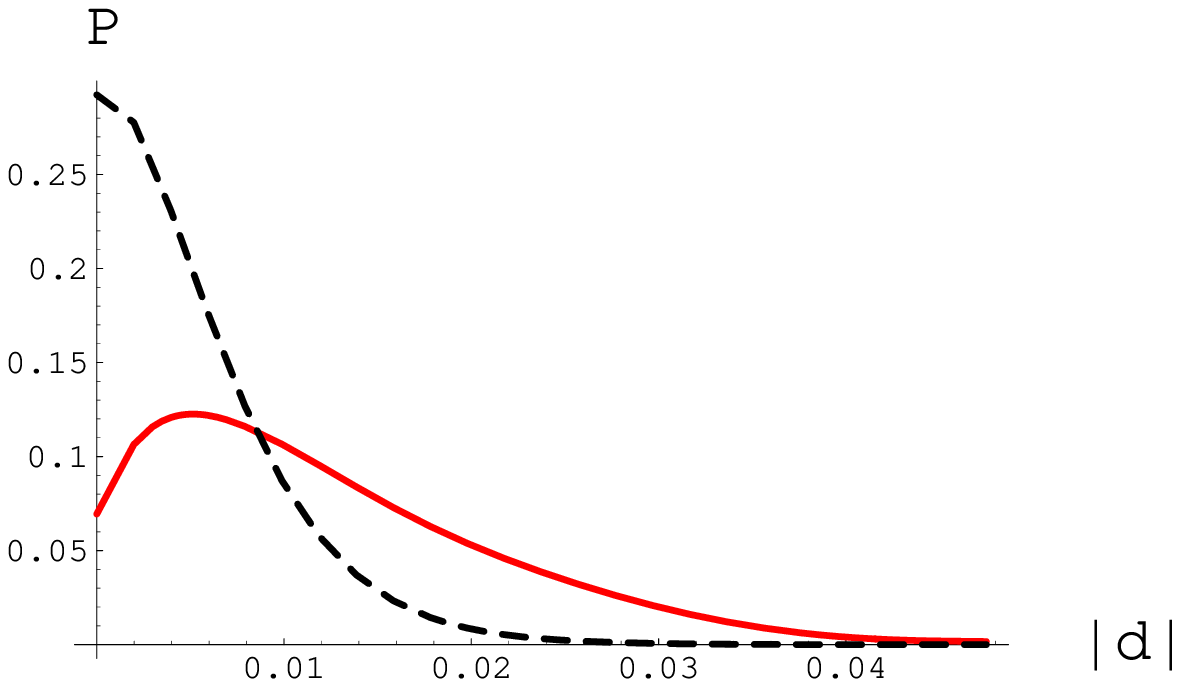}} \caption[mapCoeffMags]
 {\centering Distributions of the magnitude of the quadrupole coefficients ($a$, left panel), the sextupole
coefficients ($b$, center panel), and the cardioid/displacement
coefficients ($d$, right panel), of background galaxies using the
radial-fit method without PSF (dashed) or the model method with
the PSF (solid).} \label{mapCoeffMags}
\end{figure}

The distributions of the magnitudes for the quadrupole, sextupole,
and cardioid/displacement coefficients of background galaxies
$(z>0.5)$ are shown in fig. \ref{mapCoeffMags}.  The radial
spreading caused by the PSF causes the map coefficients to be
smaller then they are for the pre-PSF shape.

\section{Clumping of curved galaxies} \label{curvedClumps}

\begin{figure}[h!]
\centering {\leavevmode
\includegraphics[width=9cm,height=4cm]{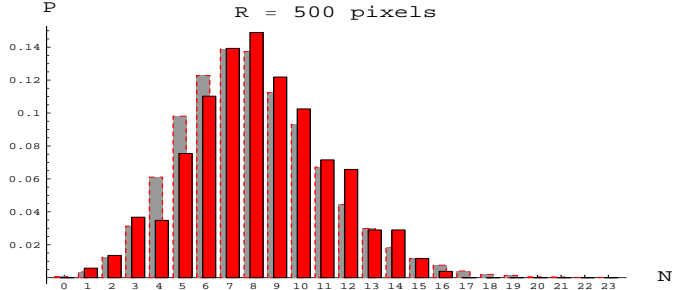}}
\caption[curvedNeighBar] {The foreground bars show the
distribution of the number of curved neighbors in circles of
radius R=500 pixels for the set of all curved galaxies. The
background shows the same distribution averaged over randomly
chosen subsets of galaxies with the same number of members. 500
pixels corresponds to 12 arcsec. Note the consistent shift to
higher numbers of neighbors.} \label{curvedNeighBar}
\end{figure}

We looked for evidence of a spatial clumping of both curved and
aligned galaxies as found in our HDFN analysis.\footnote{A
``curved" (``aligned") galaxy was defined as having a minimum
(maximum) of its sextupolar distortion within $ 10^\circ$ of one
of the two minima (maxima) of its quadrupolar distortion.} The UDF
was observed with the WFC camera which has a 2x smaller pixel size
than the WFPC used for the HDFN. Because the total exposure time
is about 10x longer, we were able to resolve 3 times more
background galaxies per sq. min.  We have used a different image
stacking process and a different, empircally based PSF
determination method. Strikingly, we find clumping of curved
galaxies in the UDF at the same spatial angular scale as found in
the HDFN. If lensing is the source of the clumping, it follows
that the scattering lenses are similar in size and mass
distribution.  The probability of the observed clumping being
random in the UDF is found to be $\approx 1\%$, as compared to our
result of $4\%$ in the HDF. The distribution of the number of
neighbors in circles of radius R=500 pixels is shown in
fig.\ref{curvedNeighBar}.  $1\%$ is the probability that randomly
chosen subsets (of all background galaxies) with the same number
of galaxies as the curved set have as many galaxies with N=8 or
more neighbors as the curved set.

The clumping we observed for aligned galaxies in the HDFN is very
weak in the UDF.

\section{Correlations in background galaxy shapes} \label{correlation}
For our UDF images we have been able to measure the magnitude and
orientation of an additional 2nd-order lensing-distortion-map
coefficient, which we have referred to as the
cardioid/displacement coefficient, or $d$-term. The $d$-term
distorts the variable $r_S^2=w_S \bar w_S$ in the argument of an
azimuthally-symmetric light stream profile with an amplitude
proportional to $r^3 \cos (\theta-\theta_d)$. To our knowledge
this is the first measurement of this coefficient with sufficient
precision to exhibit the fundamental correlation we now describe.

\begin{figure}[h!]
\centering {\leavevmode
\includegraphics[width=5.5cm,height=4cm]{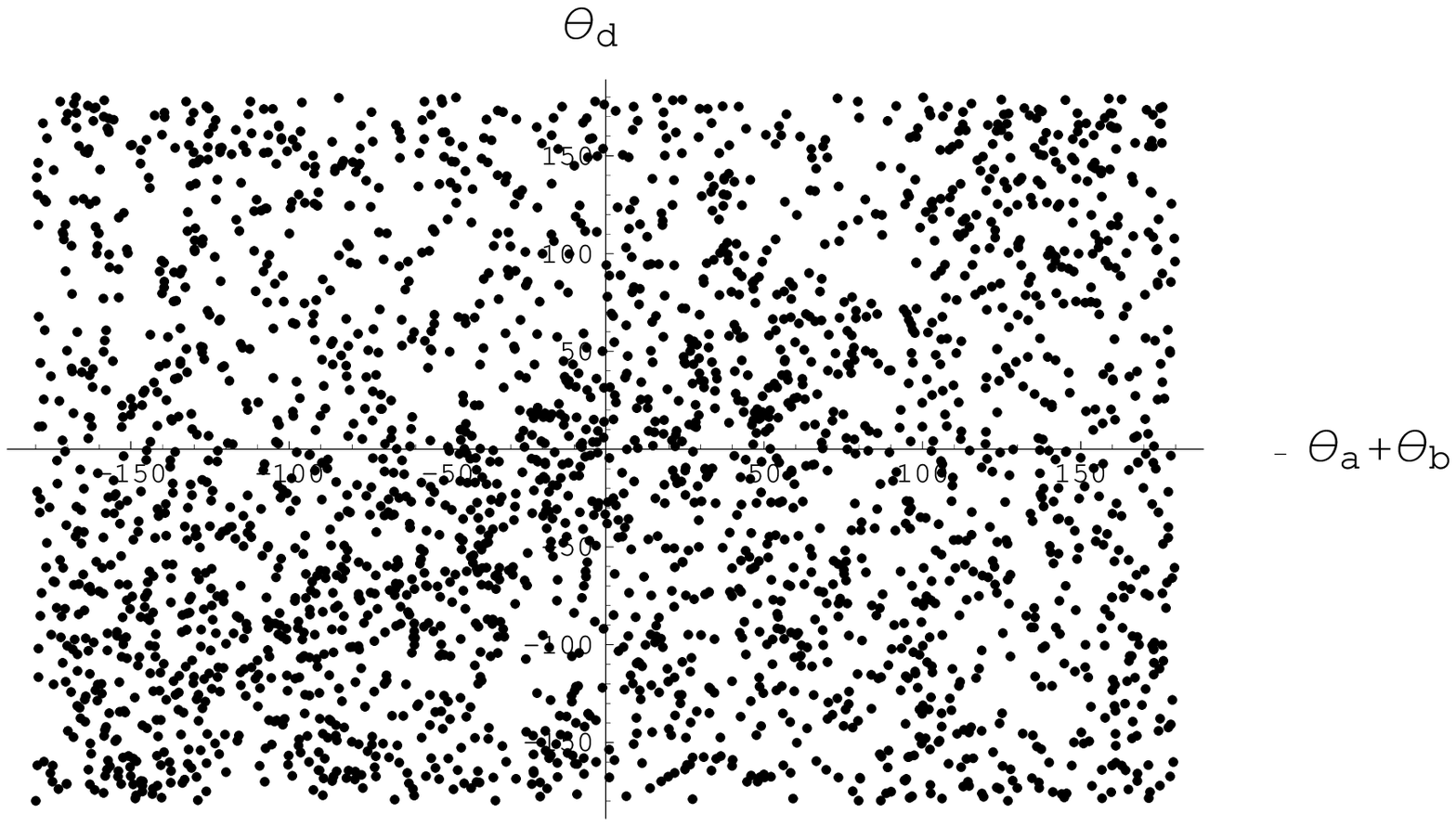}\,\,
\includegraphics[width=5.5cm,height=4cm]{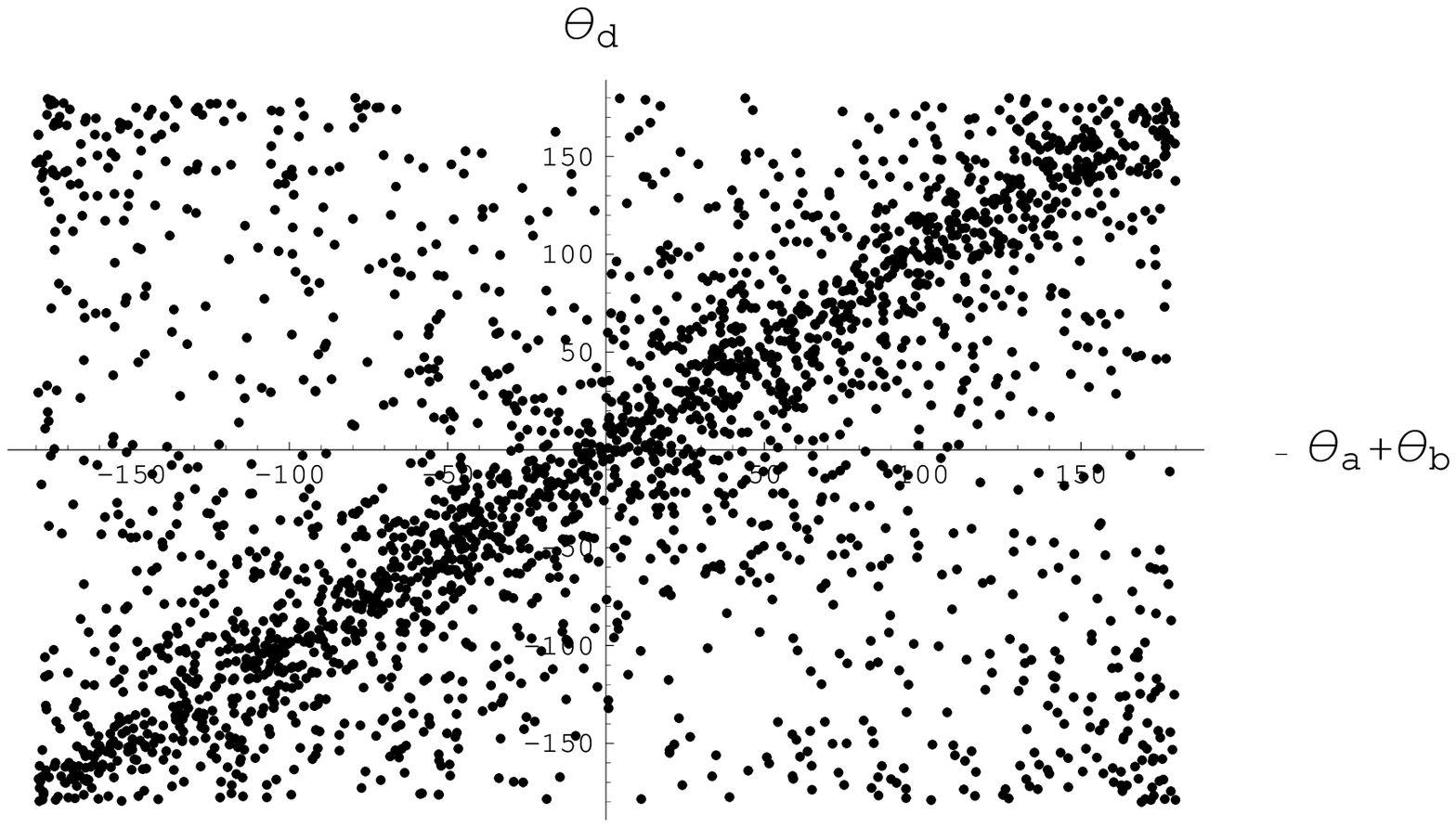}\, \,
\includegraphics[width=5.5cm,height=4cm]{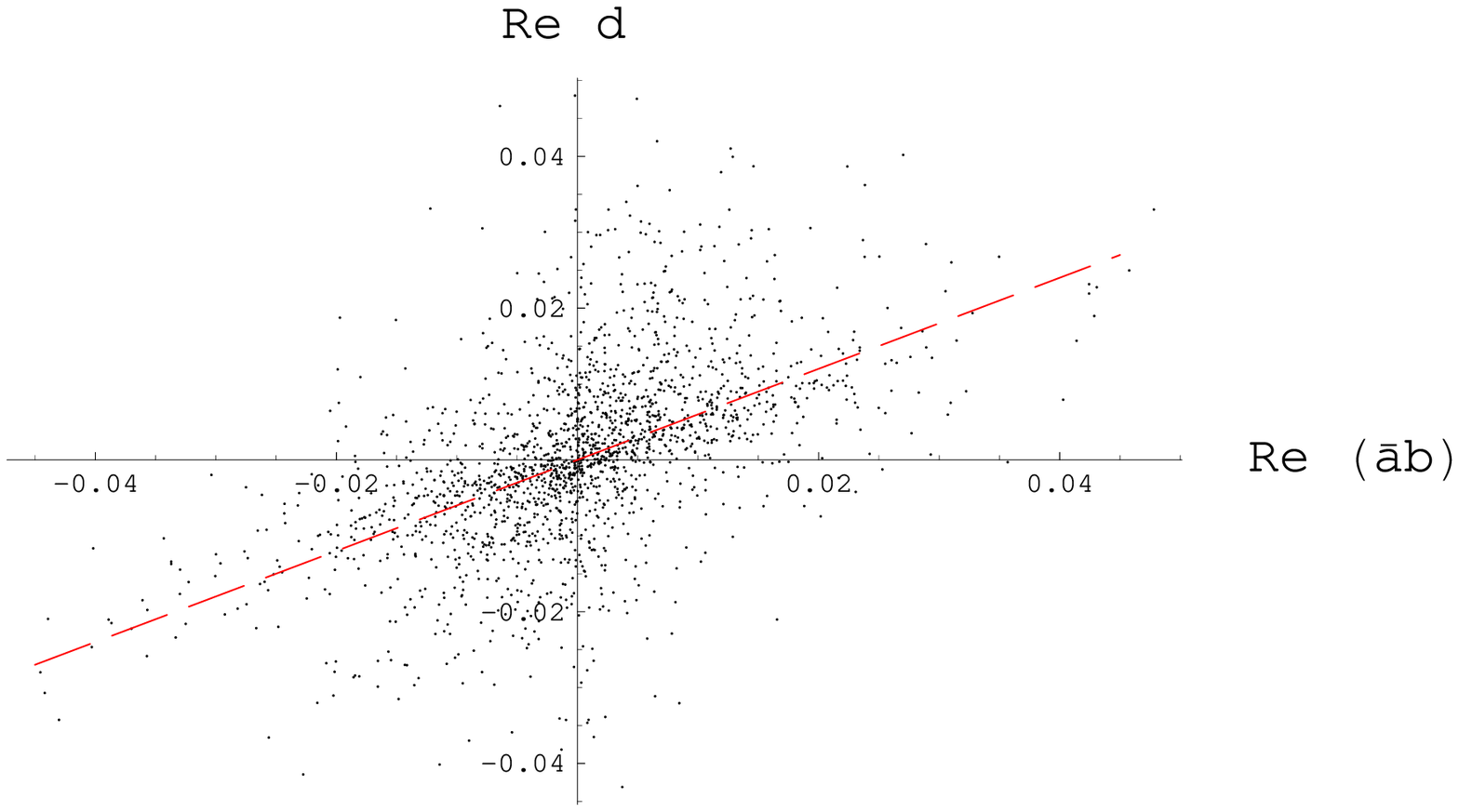}}
\caption[dAngleCorrelation] {\centering The angle of the d-term
($\theta_d$) vs. the angle of $\bar a b$ ($\theta_b-\theta_a$),
without including the PSF (left panel) and including the PSF
(center panel).  The right plot shows $Re[d]$ vs. $Re[\bar a b]$.}
\label{dAngleCorrelation}
\end{figure}

We find the orientation of the $d$-term coefficient ($\theta_d$)
to be correlated with the $\bar a b$ orientation ($\theta_b
-\theta_a$). Scatter plots of $\theta_d$ vs. ($\theta_b
-\theta_a$) are shown in the left and center panels of fig.
\ref{dAngleCorrelation}. The center plot, in which the correlation
is obvious, includes the PSF during the fit whereas the left plot,
which barely has an indication of the correlation, ignores the
PSF, illustrating the significance of the accuracy of the PSF
treatment. We have studied the properties of the PSFs and find no
evidence there for this correlation.

\begin{figure}[h!]
\centering {\leavevmode
\includegraphics[width=5.8cm,height=4.5cm]{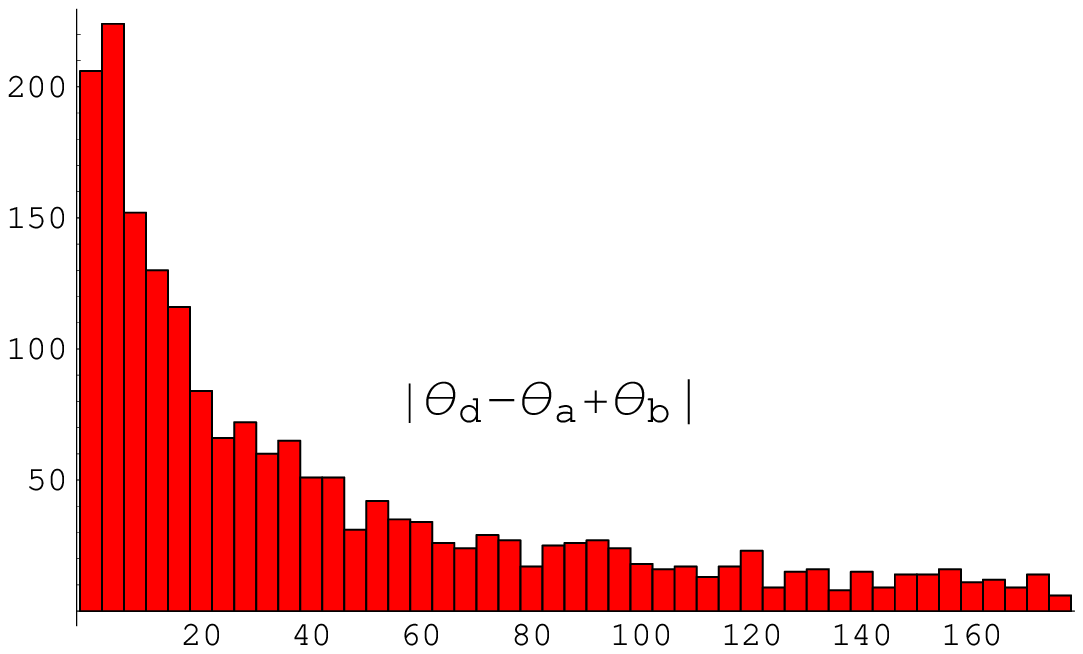}\,\,\,
\includegraphics[width=5.8cm,height=4.5cm]{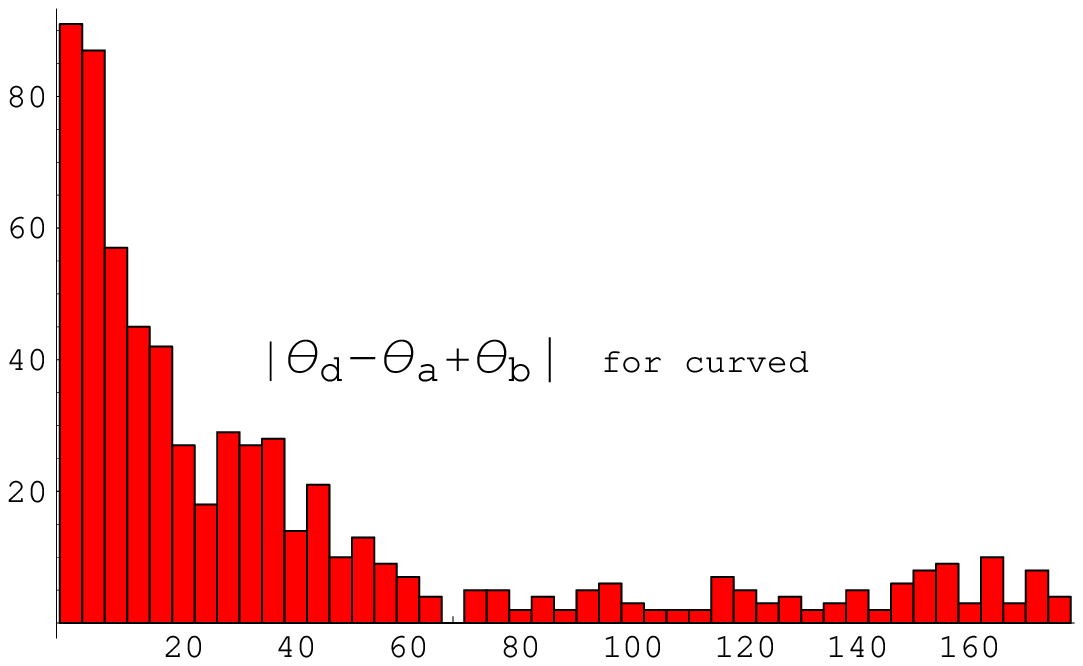}\,\,\,
\includegraphics[width=5cm,height=4.2cm]{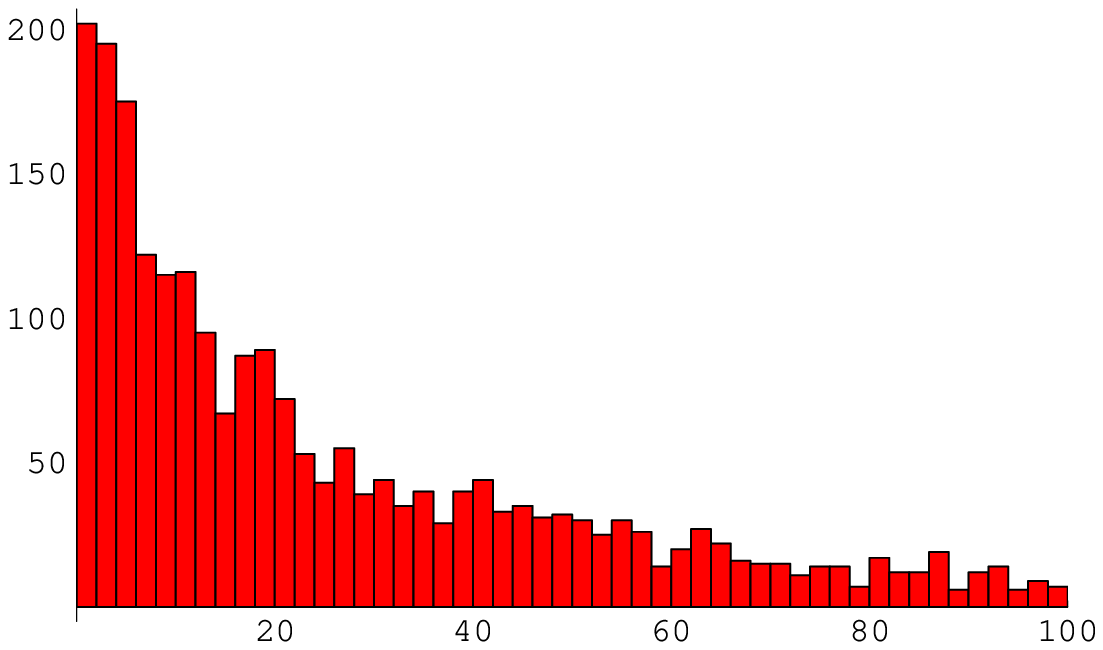}}
\caption[dAngleDistribution] {\centering The left (center) panel
shows the distribution of the angular difference $\vert \theta_d -
(\theta_b - \theta_a)\vert$ mod $180^\circ$ for all (curved)
galaxies. The right panel shows our estimate, based on Fisher
matrix calculations for each galaxy, of the expected measurement
spread in  $\vert \theta_d - (\theta_b - \theta_a)\vert$ for a
$\delta$-function correlation.} \label{dAngleDistribution}
\end{figure}

A projected distribution of galaxy count vs. $\theta_d - (\theta_b
- \theta_a)$ is shown for all galaxies in the left panel of fig.
\ref{dAngleDistribution}, and for curved galaxies in the center
panel.  For comparison, in the right panel we show the shape that
would be predicted for an underlying delta-function correlation
that incorporates the noise as predicted from a Fisher matrix
study of the determination of these angles.  The widths are
similar, suggesting that the underlying correlation, though not
necessarily a delta function, is indeed considerably sharper.

As the reference frame of the galaxy is rotated, the angles
$\theta_d$ and $\theta_b-\theta_a$ increase linearly, explaining
why the correlation in the center panel of
fig.\ref{dAngleCorrelation} lies along a line of unit slope, and
making it evident that this correlation is invariant under
rotation. A feature of the PSF or an effect such as CTE (charge
transfer efficiency), would not possess rotational invariance, and
thus could not lead to such a correlation.

Another perspective on this correlation can be seen by plotting
the real (imaginary) part of $d$ versus the real (imaginary) part
of $\bar a b$. This is done in the right panel of fig.
\ref{dAngleCorrelation}. One sees a general tilt to this diagram
along the line $d=\frac{3}{5} \bar a b$. The factor $3/5$ has been
determined, to a few percent, by a least-squares fit of a line to
the data.

As a guide to find further correlations, one can use the moment
methods of section 3.2 in [IS] to find the relationship $d -
\frac{4}{5} a \bar d = \frac{1}{5} \bar a b - \frac{1}{5}(1-\vert
a \vert^2) m_{21}$ where $m_{21} \equiv \frac {M_{21}}{M_{22}}$
and $M_{nm}=\int w^n \bar w^m i dA$. This equation was derived
assuming an azimuthally symmetric root galaxy was mapped to fit a
background shape. $M_{21}$ and $M_{22}$ are moments of the image
of the root galaxy after the mapping.

An equation of this form is to be expected from general
considerations. The quantities $d$, $b$, and $m_{21}$ are small
enough that they occur only in first order. Furthermore as the
reference coordinate system is rotated, $d$, $\bar a b$, $m_{21}$,
$a^2 \bar b$, $a \bar d$, and $a \bar m_{21}$ all rotate as
vectors, so only these quantities can be involved.  The terms with
$m_{21}$ and $a \bar m_{21}$ can be added together and treated as
the unknown. And only one of the two quantities $a \bar d$ and
$a^2 \bar b$ need be retained, as they are redundant.

Minimizing the (combined) $m_{21}$ term results in the approximate
equation \BA \label{dEquation} d - \frac{4}{5} a \bar d \approx
\frac{3}{5}
 \bar a b - \frac{1}{5}(1-\vert a \vert^2) m_{21}. \EA

\noindent The $3/5 \, \bar a b$ and $4/5 \, a \bar d$ are good
fits to the data, and an additional $a^2 \bar b$ term, if present,
has a very small coefficient.  The $1-\vert a \vert^2$ dependence
of the coefficient of $m_{21}$ was adopted from the moment
equation.

\section{Evidence for cardioid/displacement lensing} \label{lensing}

The utility of the lensing-distortion-map parameterization is that
if a background galaxy is subsequently lensed, this correlation,
expressed in terms of the observed total $a \approx a_L + a_S$ and
$b \approx b_L+b_S$, where $L$ indicates lensing and $S$ denotes
the fit to the background source galaxy, would break down even if
the lensing is quadrupole lensing alone. In other words, we know
that in the presence of lensing, the map coefficients representing
the shape of the source galaxy will add linearly to the lensing
map coefficients.  Since a contribution from lensing can be
expected to change these coefficients, lensing will disturb this
correlation condition.  This will show up as a larger magnitude of
$m_{21}$ when it is calculated using eq.\ref{dEquation}, because
the change in $3/5 \, \bar a b$ will be vectorially added to the
existing $m_{21}$.

\begin{figure}[h!]
\centering {\leavevmode
\includegraphics[width=5.5cm,height=4cm]{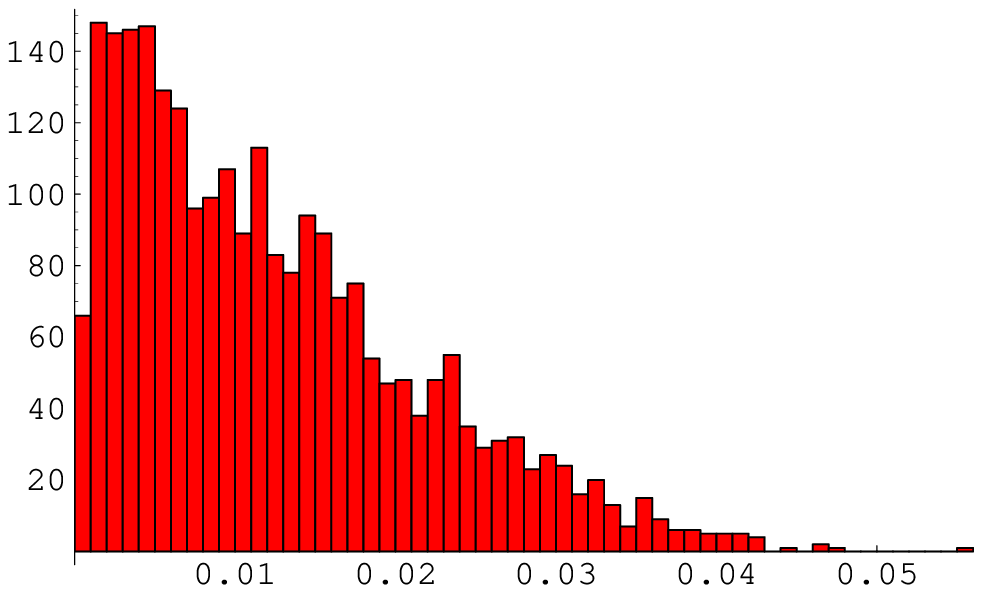}\,\,\,\,\,
\includegraphics[width=5.5cm,height=4cm]{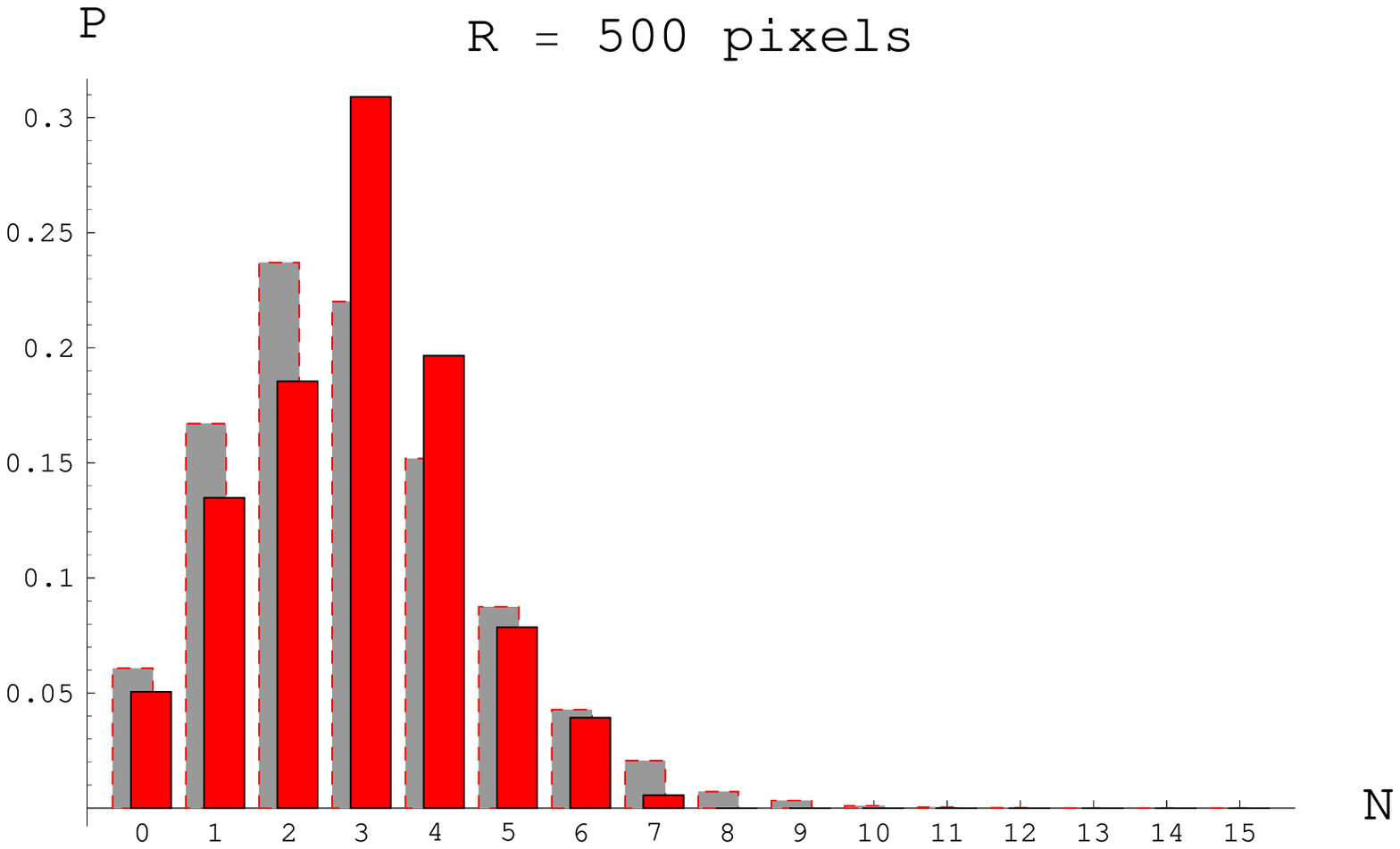}\,\,\,\,\,
\includegraphics[width=5.5cm,height=4cm]{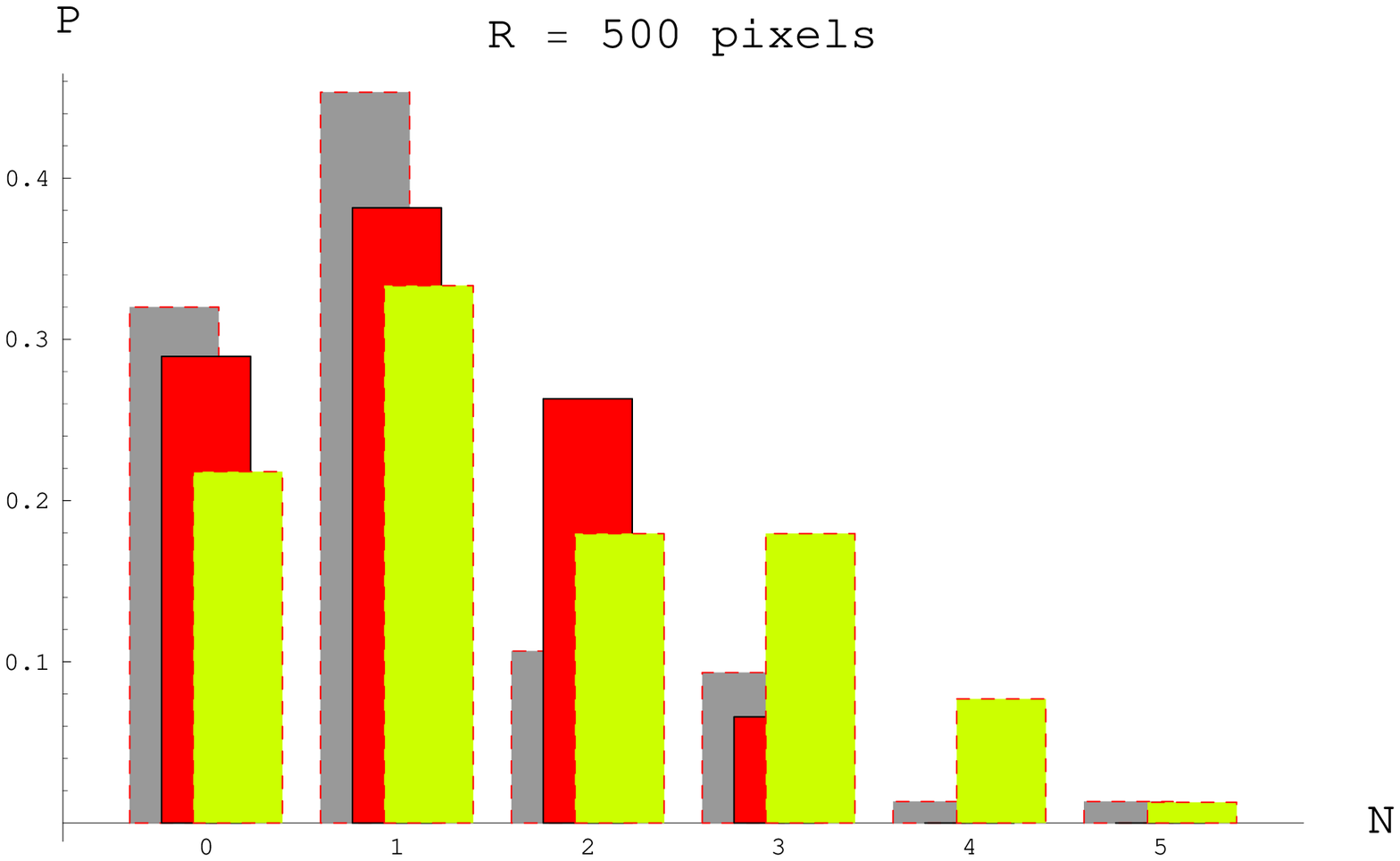}}
\caption[dAngle180] {\centering The left panel shows the
distribution of the magnitudes of $m_{21}$ as calculated from
eq.\ref{dEquation}.  Large $\vert m_{21} \vert$ is a signal for a
violation of the $d$-term correlation. The central panel shows the
neighbor distribution for 225 galaxies with $\vert m_{21}
\vert>0.023$. The background shows the neighbor distribution
averaged over randomly chosen subsets of galaxies with 225
members. Note the consistent shift to higher numbers of neighbors.
The probability of this shift occurring randomly is $<1\%$.  The
right panel shows the distribution of the number of neighbors for
three subsets of the curved galaxy distribution with 90 members:
1) $0-4.3^\circ$ (background), 2)$45-100^\circ$, and
3)$100-180^\circ$ (foreground).} \label{dAngle180}
\end{figure}

Hence if we choose those galaxies that are in the tail of the
$\vert m_{21} \vert$ distribution (see the left panel of
fig.\ref{dAngle180}) they should be more likely to be lensed
galaxies. We chose the cut at $\vert m_{21} \vert > 0.023$ to end
up with $\approx 200$ galaxies.  Sets with less than 200 members
have poorer statistics for clumping studies. The neighbor
distribution for galaxies with $\vert m_{21} \vert>0.023$ is shown
in the central panel of fig.\ref{dAngle180}. It has a very small
probability of occurring randomly, less than $1\%$.

Additionally we have investigated the small but tantalizingly
distinct hump in the distribution of the curved galaxies shown in
fig. \ref{dAngleDistribution} near $\theta_d -(\theta_b-\theta_a)
= 180^\circ$. For these galaxies the sign of the coefficient of
$d$ is opposite to that of $\bar a b$. This is the orientation for
lensing events that are non-zero for all three lensing
coefficients, and for which the $d$-term coefficient would be
responding to the presence of $\Sigma'$ (the gradient of the
projected mass density function) in the lensing plane.

Since this is a small set of galaxies, it is difficult to
convincingly establish lensing as a probable cause of this hump by
a study of clumping probabilities. However we have done that, and
when compared to randomly chosen curved galaxies of the same size,
it has a probability of arising randomly of $\approx 3 \%$.  It is
perhaps more convincing to look directly at the neighbors
distribution, as shown in the right panel of fig.\ref{dAngle180}.
The neighbors distribution is compared for three distinct subsets
of curved galaxies having the same number of members.  The galaxy
subset with $150^\circ< \theta_d-(\theta_b-\theta_a) <180^\circ$
clearly has larger numbers of neighbors. Though this is
suggestive, we would not claim to have conclusively observed
$\Sigma'$ lensing.

\section{Conclusions} \label{conclusions}
We have confirmed in the UDF the spatial clumping of curved
galaxies observed by [IS] in the \emph{HST} Deep Field North. The
clumping radius is the same as found in the HDFN with a smaller
(now $\sim 1\%$) probability of occuring randomly for the UDF. We
are now confident that we have observed nonlinear lensing due to
small-scale structure.

We have measured, for the first time, the cardioid/displacement
map coefficient ($d$-term) and its correlation with the quadrupole
and sextupole.  This correlation can be explained, in the absence
of lensing, as arising from an $M_{30}^S$ moment in the source
galaxies, which gives rise to a $b$-term and also, through $\bar a
b$, to a $d$-term in the map. Galaxies not exhibiting this
correlation are expected to be lensed, and we have presented
evidence for the spatial clumping of the set of 200 galaxies
having the largest values for the magnitude of $m_{21}$.  The
probability of this clumping being random was $<1\%$.

We have also used the orientation of the $d$-term with respect to
the direction of $\bar a b$ to identify a set of about 80 galaxies
that could have been lensed by the density derivative, $\Sigma'$.
As would be expected from strongly lensed galaxies, this subset
consists almost exclusively of curved galaxies and appears to be
spatially clumped.

As larger fields are observed, we believe there will arise the
opportunity to accurately quantify the spatial structure and
strength of small impact-parameter lensing events. We would expect
the study of these events to provide a cosmological signal of
small-scale substructure, and perhaps even be able to measure the
extent to which sub-haloes are stripped.



\newpage

\begin{thebibliography}{}

\bibitem[Anderson and King(2006)]{Anderson:2006}
J.~Anderson and I.R.~King, STSCI Instrument Science Report
ACS/ISR-2006-01.

\bibitem[Anderson(2005)]{Anderson:2005}
J.~Anderson, The 2005 HST Calibration Workshop, STSCI, 2005, p11

\bibitem[Bacon et al.(2005)]{Bacon:2005qr}
  D.~J.~Bacon, D.~M.~Goldberg, B.~T.~P.~Rowe and A.~N.~Taylor,
  arXiv:astro-ph/0504478.

\bibitem[Bertin and Arnouts (1996)]{sextractor}
E.~Bertin and S.~Arnouts, Astron. Astrophys. Suppl. Ser.{\bf 117},
393 (1996)


\bibitem[Goldberg, and Bacon (2005)]{Goldberg:2004hh}
  D.~M.~Goldberg and D.~J.~Bacon,
  Astrophys.\ J.\  {\bf 619}, 741 (2005)
  [arXiv:astro-ph/0406376].

\bibitem[Irwin and Shmakova (2003b)]{Irwin:2003qw}
J.~Irwin and M.~Shmakova,
arXiv:astro-ph/0308007.

\bibitem[Irwin and Shmakova (2005)]{Irwin:2005nc}
  J.~Irwin and M.~Shmakova,
  2006 ApJ 645,1 [arXiv:astro-ph/0504200].

\bibitem[Tiny Tim (2004)]{TinyTim:2004}
J. Krist, R. Hook, 2004
  http://www.stsci.edu/software/tinytim.



\end{thebibliography}
\end{document}